\documentstyle[12pt]{article}
\begin{document}

\newcommand{\be}{ \begin{equation}}
\newcommand{\ee}{ \end{equation}  }
\newcommand{\bi}{\bibitem}
\newcommand{\s}{ \sigma }
\newcommand{\rag}{ \rangle }
\newcommand{\lag}{ \langle }
\newcommand{\rd}{ \mbox{\rm d} }
\newcommand{\rD}{ \mbox{\rm D} }
\newcommand{\re}{ \mbox{\rm e} }

\renewcommand{\thefootnote}{\fnsymbol{footnote}}

\renewcommand{\floatpagefraction}{0.8}
\begin{titlepage}
\begin{flushright} CERN-TH 6636/92 \end{flushright}
\vspace{1ex}
\vspace{0.75cm}
\begin{center}
{ \LARGE High Precision Simulation Techniques \\
for Lattice Field Theory
\footnote[2]{Talk given at the 4th International Conference on Computational
Physics PC92,
Prague, August 1992}}
\vspace{0.75cm}

{\large Ulli Wolff \\
CERN, Theory Division,\\
CH-1211 Gen\`eve 23, Switzerland}
\date{}
\end{center}
\vspace{1cm}
\thispagestyle{empty}
\begin{abstract}\normalsize
An overview is given over the recently developed and now widely used
Monte Carlo algorithms with reduced or eliminated critical slowing down.
The basic techniques are overrelaxation, cluster algorithms and
multigrid methods. With these tools one is able to probe much closer than
before the universal continuum behavior of field theories on the lattice.
\end{abstract}
\vspace{2.cm}
\begin{flushleft} CERN-TH 6636/92 \\ September 1992 \end{flushleft}
\vspace{1ex}
\end{titlepage}
\section{Introduction}
\noindent
The interest in statistical systems close to criticality is
shared by a large community including  condensed
matter physicists and particle theorists.  An important tool in the
study of such systems are numerical experiments in the form of Monte
Carlo simulations which complement analytical results that are
available for special systems and limiting cases. One of the
central problems in such simulations is the degradation
of most known simulation algorithms as criticality --- the cutoff or
continuum limit in quantum field theories describing particles ---
is approached.  The interdisciplinary search of improved techniques
is an on-going effort, but it already has yielded some very positive
results in recent years some of which we want to briefly review here.
Our focus will be on spin systems with variables belonging
to continuous manifolds ($\s$-models) and on pure lattice gauge theory.
This selection thus leaves out the vast field of discrete Ising-like
systems$^1$
as well as the enormous problems (and potential gains from algorithmic
improvement)
faced in QCD simulations with fermions.$^2$

The problem of critical slowing down (CSD)
near criticality is schematically described by the dynamical
scaling law
\be \label{z}
\tau = c \; \xi^z.
\ee
Here $\xi$ is some physical scale or correlation length in lattice units
and $\tau$ is a time scale in number of iterations.
A $\tau$ and corresponding $z$ can characterize
the time scale for equilibration or
the rate at which
statistically independent estimates for observables of interest are
produced. Then for the expectation value of $A$,
\be \label{avA}
\lag A[s] \rag = \frac{1}{Z}\int \rD s \; \re^{-\beta H[s(x)]} \; A[s],
\ee
the accuracy in estimating $A$ improves as
\be
\delta_A =  \sqrt{\mbox{variance}_A \times 2\tau/\#\mbox{ of iterations}}\;.
\ee
In (\ref{avA}) $\rD s$ = $\prod_x \rd\mu(s(x))$ means integration
over all fields or spins $s(x)$ at lattice sites $x$
with some measure, $Z$ is the partition function and
$\beta H[s]$ is the inverse bare coupling or temperature times
some action or Hamiltonian.

The algorithms to be described here have been designed to lower $z$
from its ``traditional'' value of about two for standard local Metropolis
methods to $z \simeq 1$ or even $z \simeq 0$ in some cases.
They thus improve the efficiency
of simulations by {\it one to two powers} of $\xi$.

Beyond the introduction this article will continue with a section on
the technique of embedded variables (common to many algorithms),
and short descriptions of cluster algorithms, multigrid methods and
overrelaxation followed by some conclusions.

\section{Updating Embedded Variables}
\noindent
Let us imagine some group of transformations $T\in G$ that
act on the configurations $s \hat{=} \{s(x)\}$,
\be
s \rightarrow T s,
\ee
such that the measure is left invariant, $\rD s = \rD T s$.
Such transformations can be local, $T s = \{ t(x) s(x) \}$,
and then $t(x)$ is a field similar to $s$ itself with values
in local group factors that make up $G$. One may however also
consider more global changes of $s$, where $T$ acts for instance on some
cluster of spins from $s$ with one and the same rotation.
For a given configuration $s$ we can now think of a statistical system
with configurations $T\in G$ and an effective or induced Boltzmann
factor $\exp(-\beta H[T s])$ whose simulation can also be considered.
Assume now that we have a Monte Carlo algorithm for this system
which is characterized by transition probabilities $w(s;T_1 \to T_2)$
that preserve $\exp(-\beta H[Ts])$ and
depend on s only through the effective Boltzmann factor such that
 $w(Ts;T_1\to T_2)$ = $w(s;T_1 T \to T_2 T)$ holds.
Then, if ervalid algorithm for the original field $s(x)$:
\begin{itemize}
\item for momentarily fixed $s=s_1$ put $T_1=Id$ as initial configuration,
\item update $T_1\to T_2$ with the $w$--algorithm,
\item take $s_2=T_2s_1$ as a new $s$.
\end{itemize}

To prove this assertion we first note that the overall transition
probability is given by
\footnote[2]{For simplicity we take a discrete group $G$ here;
in the continuous case the sum over all elements has to be replaced by
an invariant group integration.}
\be
W(s_1 \to s_2) = \sum_T w(s;Id \to T) \; \delta(s_2-Ts_1).
\ee
We have to show that $W$ preserves the Boltzmann weight $\exp(-\beta H[s])$.
To this end we transform
\begin{eqnarray}
\int \rD s_1 \; \re^{-\beta H[s_1]} \; W(s_1 \to s_2) &=& \nonumber\\
\int \rD s_1 \sum_{T'} \re^{-\beta H[s_1]} \;
 w(s_1;Id \to T') \; \delta(s_2-T's_1) \; &=& \nonumber\\
\frac{1}{|G|}\int \rD s_1 \sum_{T,T'} \re^{-\beta H[Ts_1]} \;
 w(s_1;T \to T') \; \delta(s_2-T's_1) \; &=& \nonumber\\
\frac{1}{|G|}\int \rD s_1 \sum_{T'} \re^{-\beta H[T's_1]} \;
 \delta(s_2-T's_1) \; &=&
\re^{-\beta H[s_2]}.
\end{eqnarray}
To arrive at the third line  changes of variables $s_1\to Ts_1$
and $T' \to T'T^{-1}$ are carried out and $T$ is averaged over.
In the last line  we absorbed $T'$ into $s_1$ and the $\delta$-function
is used.

In many cases,
in particular if $G$ is a lower dimensional manifold than the original
configuration space, the moves described are not ergodic. This can often
be improved by using families of different embeddings between which
one switches
in a random or deterministic order. In cases where this is not sufficient
one can always blend in conventional update steps to achieve ergodicity.

\section{Cluster Algorithms}
\noindent
Cluster algorithms$^3$ for continuous spins
are an example of the successful use ovariables. Their drawback has been up to
 now that
the $O(n)$ invariant
$n$-vector models are the only continuous variable $\s$-models
where they are powerful.\footnote[2]{There are principal
reasons for this limitation which
come close to a no-go theorem.$^4$}
Here however, according to accumulated
numerical evidence$^5$, they really achieve $z\simeq 0$ with even
a very small coefficient $c$ in (\ref{z}). With the additional advantage
of variance reduced estimators for Green functions, these models have
become an ideal testing ground for nonperturbative physics$^6$,
in particular
in two dimensions, where they are asymptotically free for $n\ge 3$.
Also the xy-model ($n=2$) is of great interest$^7$ as it allows for
checks of the Kosterlitz Thouless scenario.
We therefore now specialize our general setup, Eq.(\ref{avA}), to
the $O(n)$ models, where
\be
s(x)\in R^n, \quad \rD s= \prod_x \rd^n s(x) \; \delta(s(x)^2-1), \quad
-H[s] = \sum_{<xy>} s(x) \cdot s(y).
\ee

The key to efficient cluster algorithms is the embedding of Ising
spins and the use of global update techniques for them.
A family of embeddings is labelled by an $n$-component unit vector
$r$ of the same type as the local spins. The group involved
is $G_r \simeq Z_2^{\# \mbox{sites}}$ corresponding to one
Ising spin $\s(x)=\pm 1$ on each site, $T \leftrightarrow \{\s(x)\}$.
They act as reflections,
\be \label{r}
s'=Ts, \quad s'(x)=s_{\perp}(x) + \s(x) s_{\|}(x),
\ee
where $\perp$ and  $\|$ refer to the $r$-direction.
The induced Hamiltonian
\be
-H[Ts] = \sum_{<xy>} r\cdot s(x) \; r\cdot s(y) \; \s(x) \s(y)
 +\mbox{terms independent of } \s
\ee
is recognized to describe a {\it ferromagnetic} Ising model with random
bond strengths $J_{<xy>} = r\cdot s(x) \; r\cdot s(y)$. When they are
multiplied along closed loops the result is always positive or zero
due to their factorized form, which shows the absence of frustration.
We also note that there is no magnetic field coupling to $\s$.
This can be regarded as being due to the reflections in (\ref{r})
being part of the $O(n)$ global symmetry which leads to a global
$Z_2$ invariance for $\s$.
For these embedded random systems the Swendsen-Wang algorithm$^8$
or its single cluster variant$^3$ work very well, actually better
than in the standard Ising model where some remaining CSD is still
detectable.

The algorithm just described is ergodic if $r$ is chosen at random
such that all directions can appear. In practical realizations it is
usually convenient to always take $r=(1,0,\ldots,0)$. Then
the embedding requires fewer operations, and ergodicity is restored
by globally $O(n)$-transforming the whole configuration with a random
rotation or reflection after a certain number of updates.

\section{Multigrid Techniques}
\noindent
Multigrid Monte Carlo (MGMC) techniques have been proposed$^9$ to
eliminate CSD by allowing for efficient moves of a critical
system on all scales. They work well on nearly gaussian systems$^{10}$ as
they do in the related problem of linear difference equations
where the method has made its first appearance.
Here we will not review the truly recursive  MGMC$^{11}$
approach but a simpler unigrid variant that was proposed recently.$^{12}$
It has become the optimal method for $\s$-models other than the
$O(n)$ family.

We now present the method for the $O(3)$-model.$^{12}$
The actual updates are performed here (and in realizations for
other $\s$-models) on embedded $U(1)$ spins of the xy-model type.
To an $O(3)$ generator corresponding to some rotation, as for
example
\be
i\lambda = \left( \begin{array}{ccc} 0 & 1 & 0 \\ -1~ & 0 & 0 \\
           0 & 0 & 0 \end{array} \right) \; ,
\ee
we couple local angles $\alpha(x)$ by
\be
Ts \hat{=} \left\{ \re^{-i\alpha(x) \lambda} s(x) \right\}.
\ee
This induces a Hamiltonian for $\alpha(x)$,
\be
-H[Ts]= \sum_{<xy>} \mbox{Re} \left( J_{<xy>} \re^{i(\alpha(x)-\alpha(y)}
\right) +\mbox{terms independent of } \alpha
\ee
with complex bond strengths
\be
J_{<xy>} = (s^1(x)+i s^2(x)) (s^1(y)-i s^2(y))=J_{<yx>}^{\ast},
\ee
where $s^a, a=1,2,3,$ denotes the components of $s$.
The random bonds generated for the embedded xy-model are again
ferromagnetic due to their factorized form. As for $U(1)$ gauge fields
their orientation has to be properly taken into account, of course.

We now need an algorithm for $\alpha(x)$. For the version of MGMC
of Ref.~12 elementary moves are performed on
$B\times B \times \ldots \times B$
subblocks of the lattice. For such blocks one has a priory fixed
profiles of kernels $K(x)$ that vanish outside the block and are smooth.
Possible choices
in arbitrary dimensions are pyramids or the lowest mode
sine waves with the block as a cavity. The kernels  appear in the
nonlocal Metropolis proposals
\be
\{\alpha(x)\} \to \{ \alpha(x) + \Psi K(x) \}.
\ee
These are accepted or rejected in the usual way, and $\Psi$ is drawn
from a symmetric distribution with a width adjusted for reasonable
acceptance. For a lattice of length $L$, which should be a power of two,
one has to hierarchically cover the lattice with blocks of sizes
$B = L/2, L/4, \ldots , 1$ with the last choice producing just local
updates. It has turned out to be important that either the superimposed
block lattice or equivalently
the field configuration is randomly translated between
updates, such that the block boundaries are not at fixed positions.
Furthermore the generator $\lambda$ is randomized to achieve ergodicity.

For $O(n)$ models this MGMC algorithm is presumably
inferior to cluster methods, although detailed comparisons would
be somewhat hardware dependent.
The importance of the technique derives however from the fact,
that it can also be used for $CP^n$-models and for $SU(n)$ valued
spins. The main change is that appropriate generators
(from $U(n)$ and $SU(n)$ respectively) have to be substituted for $\lambda$.
The resulting embedded $U(1)$ system now {\it can have frustration}
depending on the configuration of the ``background'' spins $s$.
Practical tests for the $SU(3) \times SU(3)$ chiral model and for
the $CP^3$ system have shown that these frustrations do not
seriously slow down the evolution.$^{13}$ In all three cases $z = 0.2(1)$
has been found.

\section{Overrelaxation}
\noindent
In contrast to the two previous algorithms overrelaxation (OR) achieves
an improvement (down to $z \approx 1$) with still local updates
only, and hence it is as fast or even somewhat faster per sweep
than standard algorithms.
It also immediately carries over to gauge fields.
We now present OR in its ``hybrid'' form that found many
applications recently rather than the original ``tunable'' version$^{14}$.
We consider the local update problem at site $x$ with local Boltzmann
weight
\be \label{loBo}
\re^{\beta s(x) \cdot M(x)} \quad
\mbox{where } \quad M(x) = \sum_{y=
\mbox{\scriptsize n.n. of } x} s(y),
\ee
again for an $O(n)$ model spin for illustration.
A local heatbath procedure amounts to choosing a new $s(x)$ independently
of the old one with the weight (\ref{loBo}). For OR we need additional
microcanonical steps producing transitions $s_1(x) \to s_2(x)$ such that
\begin{itemize}
\item $s_1(x), s_2(x)$ have the same local weight (\ref{loBo}) and thus
the energy is unchanged,
\item $s_1(x), s_2(x)$ are as far from each other as possible.
\end{itemize}
For our example this principle leads to the change
\be
s_2(x) = -s_1(x) + 2 \frac{M \; M\cdot s_1(x)}{M^2}.
\ee
Experience has shown that both for vectorization and for the nonuniversal
it is best to group the local updates such as to do a maximum number
of independent ones in parallel, which usually amounts to checkerboard
ordering.
Now an OR iteration consists of $N$ microcanonical sweeps followed
by one heatbath or other standard ergodic step.
{}From exact gaussian analysis$^{15}$ and from numerical experiments$^{16}$
it is known that to achieve $z\approx 1$ it is necessary do let $N$ grow
proportional to $\xi$ as criticality is approached. Often $N=\xi$
is a reasonably good trial value. The goal is to achieve a roughly constant
autocorrelation time when measured in iterations which
implies $z\approx 1$ when referring to sweeps.

In particular, this kind of OR is the present method of choice
for SU(2) lattice gauge fields
(either fundamental or embedded to move SU(3) fields).
The local problem for a link variable
$U_{\mu}(x) \in SU(2)$ coincides with the $O(4)$ case when $SU(2)$
matrices are expanded in terms of the unit- and the Pauli-matrices.

\begin{figure}
\vspace{9.5cm}
\caption{Autocorrelation time in sweeps for
four dimensional $SU(2)$ lattice gauge theory
in a finite size scaling limit.}
\end{figure}
We close with the example of
 a recent simulation of the $SU(2)$ gauge theory$^{17}$
 where it was possible to
determine the relation between $\tau$ and a scale in lattice units
for a whole range of scales as shown in Fig.~1.
In this study a renormalized coupling
constant was held fixed which is equivalent to a finite size scaling limit
at fixed $L\sqrt{K}$ with the
string tension $K$ assuming the r\^ole
of a correlation length.
The time $\tau$ refers to independently estimating the
renormalized coupling.
The line in the plot represents a fit with the
form (\ref{z}) giving $z=1.0(1)$ and $c=0.5(1)$.
For further details on algorithmic and physical aspects of this work
we have to refer to Ref.~17.
\section{Conclusions}
\noindent
We have presented some of the accelerated algorithms for the Monte
Carlo simulation of spin
and gauge theories that have been discovered and tested in recent years.
As a result, critical continuum behavior can be studied now much more
accurately, especially in two and three dimensions. In four dimensions,
which is the most interesting case for particle physics, the situation
will become similar as larger systems will be studied on future computers.
In the presence of Goldstones modes, the new techniques are crucial
already now.

When algorithms  of the multigrid type with $z < 1$ will be found for
gauge theories, it will strongly depend on the overhead inflicted,
at which system size they really become profitable.
In simulations of the $CP^3$ model, for instance, it has been found
that on vector machines the crossover between OR and MGMC occurs only
for correlation lengths $\xi = 20 \ldots 30$.$^{13,18}$

\vspace{1ex}

\noindent
{\large\bf Acknowledgements} \newline
I would like to thank Martin Hasenbusch and Steffen Meyer for
correspondence and discussions on their multigrid technique.

\end{document}